\documentclass[10pt]{article}
\usepackage{multicol}
\usepackage{graphicx}
\usepackage{amsmath}
\usepackage[a4paper]{geometry}
\usepackage{rotating}

\begin{document}
\begin{center}
{\Large \bf Energy dependent kinetic freeze-out temperature and\\
transverse flow velocity in high energy collisions}

\vskip.75cm

Li-Li Li, Fu-Hu Liu{\footnote{E-mail: fuhuliu@163.com;
fuhuliu@sxu.edu.cn}}

\vskip.25cm

{\small\it Institute of Theoretical Physics and State Key
Laboratory of Quantum Optics and Quantum Optics Devices,\\ Shanxi
University, Taiyuan, Shanxi 030006, China}
\end{center}

\vskip.5cm

{\bf Abstract:} Transverse momentum spectra of negative and
positive pions produced at mid-(pseudo)rapidity in inelastic or
non-single-diffractive proton-proton collisions and in central
nucleus-nucleus collisions over an energy range from a few GeV to
above 10 TeV are analyzed by a (two-component) blast-wave model
with Boltzmann-Gibbs statistics and with Tsallis statistics
respectively. The model results are in similarly good agreement
with the experimental data measured by a few productive
collaborations who work at the Heavy Ion Synchrotron (SIS), Super
Proton Synchrotron (SPS), Relativistic Heavy Ion Collider (RHIC),
and Large Hadron Collider (LHC), respectively. The energy
dependent kinetic freeze-out temperature and transverse flow
velocity are obtained and analyzed. Both the quantities have a
quick increase from the SIS to SPS, and slight increase or
approximate invariability from the top RHIC to LHC. Around the
energy bridge from the SPS to RHIC, the considered quantities in
proton-proton collisions obtained by the blast-wave model with
Boltzmann-Gibbs statistics show a more complex energy dependent
behavior compared with the results in the other three cases.
\\

\\


\vskip1.0cm

\begin{multicols}{2}

{\section{Introduction}}

The initial state, chemical freeze-out, and kinetic freeze-out are
three main stages undergone by the interacting system in
particle-particle, particle-nucleus, and nucleus-nucleus
collisions at high energies. As the basic element (proton and
neutron) in proton-nucleus and nucleus-nucleus collisions,
proton-proton collisions display some similarities to
proton-nucleus and nucleus-nucleus collisions. In particular,
proton-proton collisions and gold-gold (copper-copper, lead-lead)
collisions have been studied in large-scale experiments at the
Heavy Ion Synchrotron (SIS), Super Proton Synchrotron (SPS),
Relativistic Heavy Ion Collider (RHIC), and Large Hadron Collider
(LHC). A few productive collaborations have been reporting
abundant data on the particle ratios, (pseudo)rapidity spectra,
transverse momentum (mass) spectra, invariant mass spectra,
anisotropic flows, nuclear modification factor, and so on. Some
useful information related to particle production and system
evolution can be extracted from these data.

``Temperature is surely one of the central concepts in
thermodynamics and statistical mechanics" [1]. The temperature in
subatomic physics describes the excitation degree of the
interacting system. Due to the system evolution, the temperature
is expected to change at different stages. Undoubtedly, the
kinetic freeze-out temperature (the temperature at the kinetic
freeze-out stage) is less than or equal to the chemical freeze-out
temperature (the temperature at the chemical freeze-out stage) due
to the fact that the kinetic freeze-out happens posteriorly or
simultaneously compared with the chemical freeze-out. The
temperature at the initial state is the highest due to the largest
compression and density during the process of system evolution.
After the initial state, the interacting system undergoes
transverse expansion and longitudinal extension. At the same time,
some particles are emitted during the evolution process, though
most particles are emitted at the kinetic freeze-out. The kinetic
freeze-out temperature and concomitant transverse flow velocity
can be extracted from the transverse momentum spectra of
identified particles.

In order to extract the kinetic freeze-out temperature and
transverse flow velocity and to study their dependence on energy,
one can use different models to analyze the transverse momentum
spectra. These models include, but are not limited to, the
blast-wave model with Boltzmann-Gibbs statistics [2, 3] or with
Tsallis statistics [4--6] and the alternative method [7--11] with
standard distribution or with Tsallis distribution, where the
standard distribution denotes the Boltzmann, Fermi-Dirac, and
Bose-Einstein distributions. The blast-wave model can obtain
simultaneously the kinetic freeze-out temperature and transverse
flow velocity from a single formula. The alternative method needs
a few steps to obtain the two quantities, where the intercept in
the linear relation of effective temperature against rest mass is
regarded as the kinetic freeze-out temperature [7--11], and the
slope in the linear relation of mean transverse momentum against
mean moving mass (mean energy) is regarded as the transverse flow
velocity [9--11]. The effective temperature is in fact the
temperature parameter in the standard distribution in which the
contribution of flow effect is not excluded. Comparatively, the
blast-wave model is more convenient than the alternative method.

As we know, the energy dependent kinetic freeze-out temperature
and transverse flow velocity in high energy collisions is an open
question at present. We think that it is hard to conclude whether
or not there is an increase, decrease, or invariant trend with the
increase of energy. In fact, the situations on the energy
dependence of the kinetic freeze-out temperature and transverse
flow velocity in the literature are contradictory. Some of them
show an increase trend [12--20], some of them show a decrease
trend [15--20], and others show an invariant trend [12, 13] in the
energy dependent kinetic freeze-out temperature and/or transverse
flow velocity, with the increase of energy from the RHIC to LHC.
Our recent work [21] shows an increase or invariant trend. As for
the energy range from the SIS to SPS, the trend is always
incremental in different articles in the literature. Indeed, It is
necessary to do more studies on this topic and to conclude some
affirmative conclusions.

In this paper, the blast-wave model with Boltzmann-Gibbs
statistics [2, 3] and with Tsallis statistics [4] is used to
extract the kinetic freeze-out temperature and transverse flow
velocity by describing the transverse momentum spectra of negative
and positive pions produced at mid-(pseudo)rapidity in inelastic
(INEL) or non-single-diffractive (NSD) proton-proton ($pp$ or
$p$-$p$) collisions and in central gold-gold (Au-Au)
[copper-copper (Cu-Cu), lead-lead (Pb-Pb)] collisions over an
energy range from the SIS to LHC. The model results are compared
with the experimental data measured at the SIS, SPS, RHIC, and LHC
by the FOPI [22], NA61/SHINE [23], PHENIX [24, 25], STAR [26--28],
ALICE [29, 30], and CMS [31, 32] Collaborations. In some cases,
the two-component blast-wave model is used due to the fact that
the single model cannot fit the data very well.

This paper is structured as follows. The formalism and method are
described in Section 2. Results and discussion are given in
Section 3. In Section 4, we summarize our main observations and
conclusions.
\\

{\section{Formalism and method}}

Generally, the spectra in the high transverse momentum region are
contributed by the hard scattering process which is described by
the quantum chromodynamics (QCD) calculus [33--35] or the Hagedorn
function [36, 37] which is in fact an inverse power law which has
at least, in our opinion, three revisions. In refs. [38],
[39--43], and [44], the three revisions of the Hagedorn function
(inverse power law) are given respectively, though specific
nomenclatures on the revised Hagedorn functions are not mentioned
in some cases. We shall not discuss the Hagedorn function and its
revisions due to the fact that the hard scattering process has no
contribution to the kinetic freeze-out temperature and transverse
flow velocity. Instead, the soft excitation process contributes
the spectra in low transverse momentum region in which the kinetic
freeze-out temperature and transverse flow velocity can be
extracted.

In order to extract the kinetic freeze-out temperature and
transverse flow velocity, we should analyze the spectra in low
transverse momentum region. Although one can choose various
functions to describe the mentioned spectra, the blast-wave model
with Boltzmann-Gibbs statistics [2, 3] and with Tsallis statistics
[4--6] is a convenient consideration. For the spectra in high
transverse momentum region, the mentioned model is also a possible
choice with larger parameter values.

According to refs. [2, 3], in the blast-wave model with
Boltzmann-Gibbs statistics, the first component with the kinetic
freeze-out temperature, $T_1$, and transverse flow velocity,
$\beta_{T1}$, results in the probability density function of
transverse momenta, $p_T$, to be
\begin{align}
f(p_T,T_1,\beta_{T1}) =& \frac{1}{N}\frac{dN}{dp_T} = C_1 p_T m_T \int_0^R rdr \times \nonumber\\
& I_0 \bigg[\frac{p_T \sinh(\rho_1)}{T_1} \bigg] K_1
\bigg[\frac{m_T \cosh(\rho_1)}{T_1} \bigg],
\end{align}
where $C_1$ is the normalized constant, $m_T=\sqrt{p_T^2+m_0^2}$
is the transverse mass, $m_0$ is the rest mass, $r/R$ is the
relative radial position, $I_0$ and $K_1$ are the modified Bessel
functions of the first and second kinds respectively, $\rho_1=
\tanh^{-1} [\beta_1(r)]$ is the boost angle, $\beta_1(r)=
\beta_{S1}(r/R)^{n_0}$ is a self-similar flow profile,
$\beta_{S1}$ is the flow velocity on the surface, and $n_0=2$ is
used in the original form [2]. As a mean of $\beta_1(r)$,
$\beta_{T1}=(2/R^2)\int_0^R r\beta_1(r)dr = 2\beta_{S1}/(n_0+2)$.

According to ref. [4--6], in the blast-wave model with Tsallis
statistics, the first component with $T_1$ and $\beta_{T1}$
results in the probability density function of $p_T$ to be
\begin{align}
f(p_T,T_1,\beta_{T1})=& \frac{1}{N}\frac{dN}{dp_T} = C_1 p_T m_T
\int_{-\pi}^{\pi} d\phi \int_0^R rdr \times \nonumber\\
& \Big\{1+\frac{q-1}{T_1} \big[ m_T \cosh(\rho) \nonumber\\
& -p_T \sinh(\rho) \cos(\phi)\big] \Big\}^{-1/(q-1)},
\end{align}
where $q$ is an entropy index that characterizes the degree of
non-equilibrium, $\phi$ denotes the azimuthal angle, and $n_0=1$
is used in the original form [4]. It should be noted that
$f(p_T,T_1,\beta_{T1})$, $T_1$, $\beta_{T1}$, and $C_1$ in Eq. (2)
are different from those in Eq. (1), though the same symbols are
used.

The second component has the same form as the first one, but with
the kinetic freeze-out temperature, $T_2$, and transverse flow
velocity, $\beta_{T2}$. The two-component blast-wave model can be
structured as
\begin{align}
f_0(p_T) =& \frac{1}{N}\frac{dN}{dp_T}= kf(p_T,T_1,\beta_{T1}) \nonumber\\
& +(1-k)f(p_T,T_2,\beta_{T2}),
\end{align}
where $k$ denotes the contribution fraction (rate) of the first
component.

According to Hagedorn's model [36], we may also use the usual step
function to structure the two-component blast-wave model. That is
\begin{align}
f_0(p_T)=& \frac{1}{N}\frac{dN}{dp_T}=A_1 \theta(p_1-p_T) f(p_T,T_1,\beta_{T1}) \nonumber\\
& +A_2 \theta(p_T-p_1)f(p_T,T_2,\beta_{T2}),
\end{align}
where $A_1$ and $A_2$ are constants which result in the two
components to be equal to each other at $p_T=p_1$,
$\theta(p_1-p_T)=1$ (or 0) if $p_T<p_1$ (or $p_T>p_1$), and
$\theta(p_T-p_1)=1$ (or 0) if $p_T>p_1$ (or $p_T<p_1$).

Both Eqs. (3) and (4) can be used to extract the kinetic
freeze-out temperature, $T_0$, and transverse flow velocity,
$\beta_T$, in the two-component blast-wave model. The first
component is used to extract $T_0$ and $\beta_T$ due to its
contribution in low $p_T$ region by the soft process, and the
second component is not used to extract $T_0$ and $\beta_T$ due to
its contribution in high $p_T$ region by the hard process. We have
\begin{align}
T_0=T_1
\end{align}
and
\begin{align}
\beta_T=\beta_{T1}.
\end{align}
In the case of using Eq. (3) to get the parameter values of two
components, $k$ is directly given by Eq. (3). In the case of using
Eq. (4) to get the parameter values of two components, $k$ is
expressed by
\begin{align}
k=\int_0^{p_1} A_1 f(p_T,T_1,\beta_{T1})dp_T
\end{align}
due to the fact that Eq. (4) is the probability density function
which results naturally in the normalization.

There are little differences between values of $T_0$ ($\beta_T$)
extracted from Eqs. (3) and (4). Generally, $T_1$ ($\beta_{T1}$)
from Eq. (3) is less than that from Eq. (4) by $\leq5\%$, and
$T_2$ ($\beta_{T2}$) from Eq. (3) is larger than that from Eq. (4)
by $\leq5\%$. We are inclined to use Eq. (3) due to its convenient
fit in obtaining a smooth curve, though there is a decussate
region between the contributions of the two components in Eq. (3).
Oppositely, with Eq. (4) it is hard to get a smooth curve at
$p_T=p_1$, though there is no entanglement between the two
components in Eq. (4).

To describe the system and statistics clearly, we divide the
subject investigated in the present work into four cases. The case
I is the case which describes $pp$ collisions by the blast-wave
model with Boltzmann-Gibbs statistics. The case II is the case
which describes $pp$ collisions by the blast-wave model with
Tsallis statistics. The case III is the case which describes
nucleus-nucleus collisions by the blast-wave model with
Boltzmann-Gibbs statistics. And the case IV is the case which
describes nucleus-nucleus collisions by the blast-wave model with
Tsallis statistics.
\\

{\section{Results and discussion}}

The transverse momentum spectra of $\pi^-$ and $\pi^+$ produced at
mid-(pseudo)rapidity in INEL or NSD $pp$ collisions at high
energies are presented in Fig.1, where different
mid-(pseudo)rapidity ($y$ or $\eta$) intervals and collision
energies ($\sqrt{s}$) are marked in the panels. The closed (open)
symbols presented in panels (a)--(e) represent the data of $\pi^-$
($\pi^+$) measured by the NA61/SHINE [23], PHENIX [24], STAR [26],
ALICE [29], and CMS [31, 32] Collaborations, respectively, where
in panel (a) only the spectra of $\pi^-$ are available, and panel
(c) is for NSD events and the other panels are for INEL events.
The solid (dashed) curves are our results calculated by Eqs. (1)
and (3) for $\pi^-$ ($\pi^+$) (case I), and the dotted
(dot-dashed) curves are our results calculated by Eqs. (2) and (3)
for $\pi^-$ ($\pi^+$) (case II). Different forms of the spectra
are used due to different Collaborations, where $N$, $E$, $p$,
$\sigma$, and $N_{EV}$ denote the particle number, energy,
momentum, cross-section, and event number, respectively. In some
cases, different amounts marked in the panels are used to scale
the data for clarity. Panels (a$'$)--(e$'$) show the ratios of
data to fit obtained from Eqs. (1) and (3) for the case I,
corresponding to the collisions shown in the $p_T$ spectra; and
panels (a$''$)--(e$''$) show the ratios of data to fit obtained
from Eqs. (2) and (3) for the case II. The values of free
parameters ($T_1$, $T_2$, $\beta_{T1}$, $\beta_{T2}$, and $q$ if
available), normalization constant ($N_0$), $\chi^2$, and degrees
of freedom (dof) corresponding to the curves in Fig. 1 are listed
in Tables 1 and 2 for the cases I and II respectively. One can see
that Eq. (3) describes similarly well the $p_T$ spectra at
mid-(pseudo)rapidity in INEL or NSD $pp$ collisions over an energy
range from a few GeV to above 10 TeV. The main parameters ($T_0$
and $\beta_T$) show some laws in the considered energy range,
which will be discussed later.

\begin{figure*}[htbp]
\begin{center}
\includegraphics[width=16.0cm]{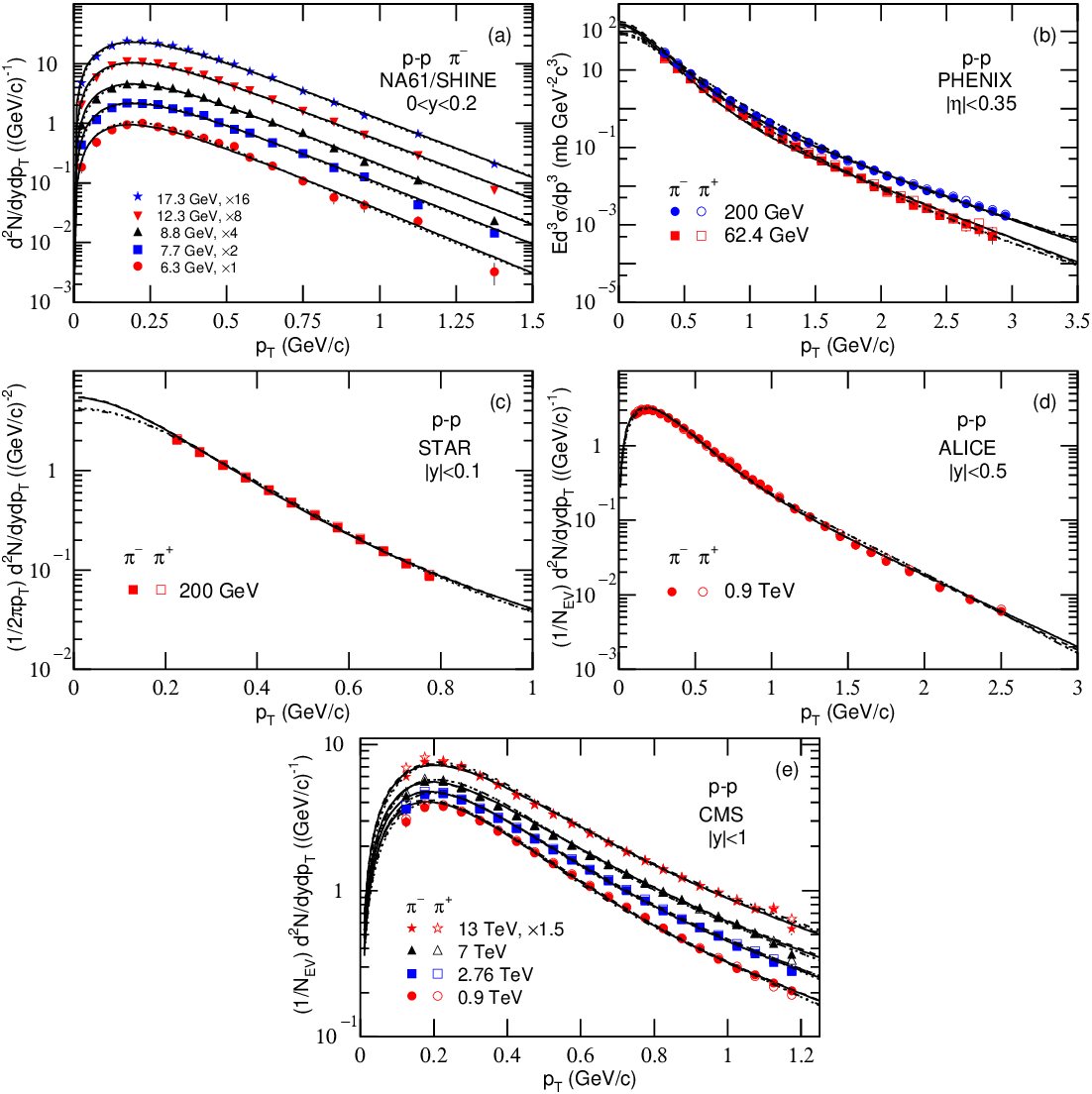}
\end{center}
{\small Fig. 1. Transverse momentum spectra of $\pi^-$ and $\pi^+$
produced at mid-(pseudo)rapidity in INEL or NSD $pp$ collisions at
high energies, where the mid-(pseudo)rapidity intervals and
energies are marked in the panels. Panels (a)--(e) represent the
data measured by the NA61/SHINE [23], PHENIX [24], STAR [26],
ALICE [29], and CMS [31, 32] Collaborations, respectively, by
various symbols, where in panel (a) only the spectra of $\pi^-$
are available, and panel (c) is for NSD events and other panels
are for INEL events. In some cases, different amounts marked in
the panels are used to scale the data for clarity. The solid
(dashed) curves are our results calculated by Eqs. (1) and (3) for
$\pi^-$ ($\pi^+$) (case I), and the dotted (dot-dashed) curves are
our results calculated by Eqs. (2) and (3) for $\pi^-$ ($\pi^+$)
(case II). Panels (a$'$)--(e$'$) and (a$''$)--(e$''$) are for
ratios of the data to fit obtained from Eqs. (1) and (3) for the
case I and from Eqs. (2) and (3) for the case II respectively,
corresponding to the collisions shown in the $p_T$ spectra.}
\end{figure*}

\begin{figure*}
\begin{center}
\vskip -1cm
\includegraphics[width=16.0cm]{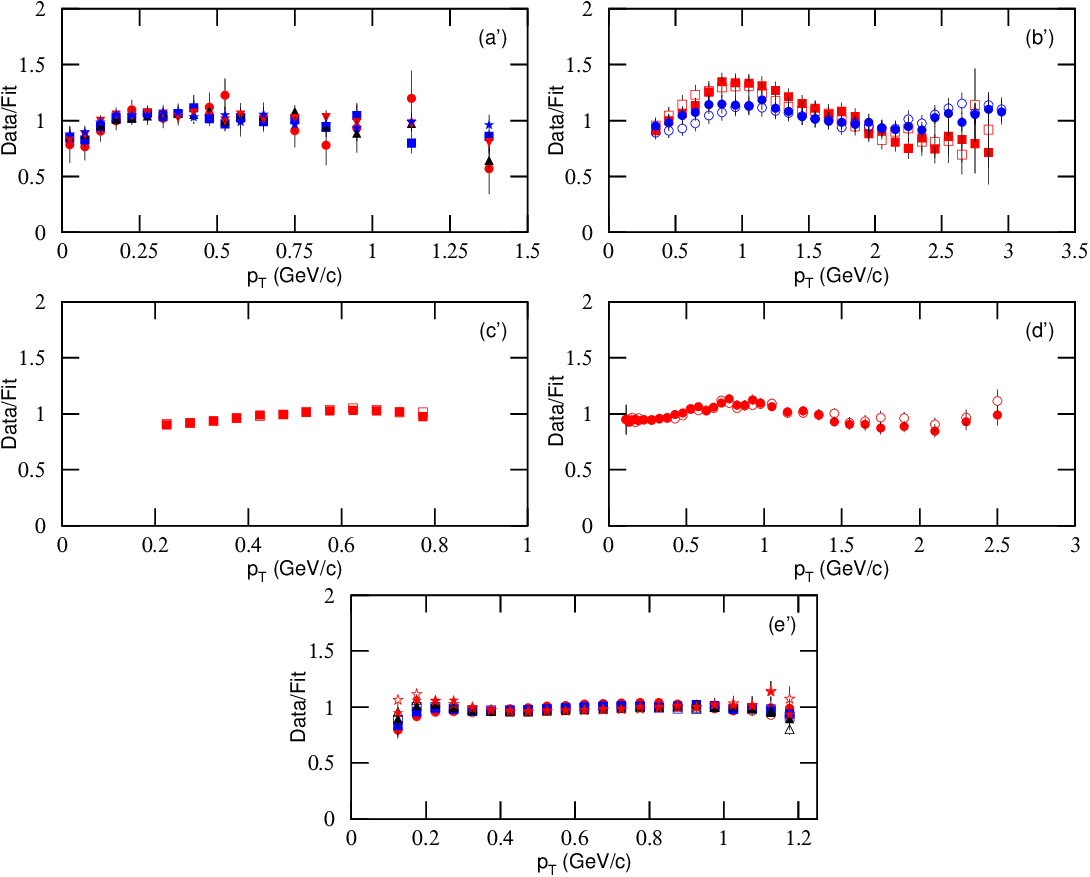}
\includegraphics[width=16.0cm]{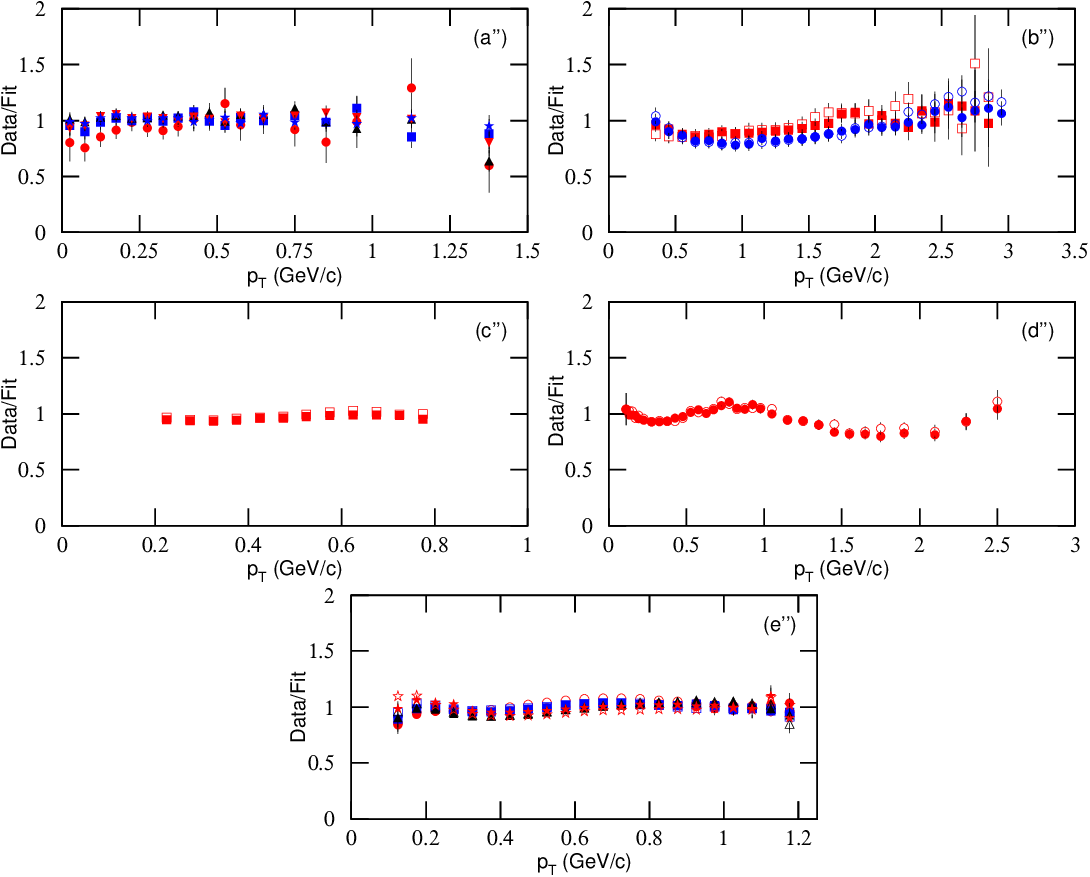}
\end{center}
{\small Fig. 1. Continued.}
\end{figure*}

\end{multicols}
\begin{sidewaystable}
%\vskip-1.cm
%\begin{table*}
{\small Table 1. Values of free parameters ($T_1$, $T_2$,
$\beta_{T1}$, and $\beta_{T2}$), normalization constant ($N_0$),
and $\chi^2$/dof corresponding to the solid and dashed curves in
Fig. 1 in which different data are measured in different
mid-(pseudo)rapidity intervals at different energies by different
Collaborations. In some cases, there is no the second component.
\begin{center}
\begin{tabular} {ccccccccc}\\ \hline\hline Collab. & $\sqrt{s}$ (GeV) &
Particle & $T_1$ (MeV) & $T_2$ (MeV) & $\beta_{T1}$ ($c$) & $\beta_{T2}$ ($c$) & $N_0$ & $\chi^2$/dof \\
\hline
NA61/SHINE & 6.3  & $\pi^-$ & $108\pm5$ & $-$        & $0.30\pm0.02$ & $-$           & $0.08\pm0.01$  &21/15\\
           & 7.7  & $\pi^-$ & $109\pm5$ & $-$        & $0.31\pm0.02$ & $-$           & $0.10\pm0.01$  &34/15\\
           & 8.8  & $\pi^-$ & $110\pm5$ & $-$        & $0.31\pm0.02$ & $-$           & $0.10\pm0.01$  &73/15\\
           & 12.3 & $\pi^-$ & $111\pm5$ & $-$        & $0.32\pm0.02$ & $-$           & $0.12\pm0.01$  &58/15\\
           & 17.3 & $\pi^+$ & $112\pm5$ & $-$        & $0.33\pm0.02$ & $-$           & $0.13\pm0.01$  &26/15\\
\hline
PHENIX     & 62.4 & $\pi^-$ & $99\pm4$  & $144\pm6$  & $0.29\pm0.01$ & $0.37\pm0.02$ & $23.04\pm1.15$ &21/20\\
           &      & $\pi^+$ & $99\pm4$  & $148\pm6$  & $0.29\pm0.01$ & $0.37\pm0.01$ & $20.81\pm1.04$ &30/20\\
           & 200  & $\pi^-$ & $103\pm4$ & $178\pm7$  & $0.29\pm0.02$ & $0.36\pm0.02$ & $27.93\pm1.38$ &8/21\\
           &      & $\pi^+$ & $103\pm4$ & $178\pm7$  & $0.29\pm0.02$ & $0.37\pm0.02$ & $27.96\pm1.56$ &17/21\\
\hline
STAR       & 200  & $\pi^-$ & $103\pm4$ & $176\pm8$  & $0.29\pm0.02$ & $0.37\pm0.02$ & $0.27\pm0.01$  &15/6\\
           &      & $\pi^+$ & $103\pm4$ & $176\pm8$  & $0.29\pm0.02$ & $0.37\pm0.02$ & $0.28\pm0.01$  &17/6\\
\hline
ALICE      & 900  & $\pi^-$ & $104\pm4$ & $200\pm9$  & $0.31\pm0.02$ & $0.36\pm0.02$ & $1.51\pm0.08$  &35/27\\
           &      & $\pi^+$ & $104\pm4$ & $203\pm9$  & $0.31\pm0.02$ & $0.36\pm0.02$ & $1.52\pm0.08$  &49/27\\
\hline
CMS        & 900  & $\pi^-$ & $104\pm4$ & $211\pm9$  & $0.32\pm0.02$ & $0.39\pm0.02$ & $3.68\pm0.19$  &12/16\\
           &      & $\pi^+$ & $104\pm4$ & $210\pm9$  & $0.32\pm0.02$ & $0.39\pm0.02$ & $3.74\pm0.20$  &10/16\\
           & 2760 & $\pi^-$ & $106\pm4$ & $232\pm8$  & $0.33\pm0.02$ & $0.39\pm0.02$ & $4.54\pm0.24$  &18/16\\
           &      & $\pi^+$ & $106\pm4$ & $231\pm9$  & $0.33\pm0.02$ & $0.39\pm0.02$ & $4.61\pm0.24$  &24/16\\
           & 7000 & $\pi^-$ & $108\pm4$ & $230\pm10$ & $0.35\pm0.02$ & $0.39\pm0.02$ & $5.61\pm0.29$  &49/16\\
           &      & $\pi^+$ & $108\pm4$ & $228\pm10$ & $0.35\pm0.02$ & $0.39\pm0.02$ & $5.74\pm0.29$  &75/16\\
           & 1300 & $\pi^-$ & $110\pm5$ & $221\pm10$ & $0.37\pm0.02$ & $0.38\pm0.02$ & $5.20\pm0.26$  &45/16\\
           &      & $\pi^+$ & $110\pm5$ & $221\pm10$ & $0.37\pm0.02$ & $0.38\pm0.02$ & $5.27\pm0.26$  &81/16\\
\hline
\end{tabular}%
\end{center}}
%\end{table*}
\end{sidewaystable}
\begin{multicols}{2}

\end{multicols}
\begin{sidewaystable}
%\vskip-1.cm
%\begin{table*}
{\small Table 2. Same as Table 1, but one more parameter ($q$) is
added and the values correspond to the dotted and dot-dashed
curves in Fig. 1.
\begin{center}
\begin{tabular} {cccccccccc}\\ \hline\hline Collab. & $\sqrt{s}$ (GeV) &
Particle & $T_1$ (MeV) & $T_2$ (MeV) & $\beta_{T1}$ ($c$) & $\beta_{T2}$ ($c$) & $q$ & $N_0$ & $\chi^2$/dof \\
\hline
NA61/SHINE & 6.3  & $\pi^-$ & $80\pm4$ & $-$        & $0.18\pm0.01$ & $-$           & $1.052\pm0.009$ & $0.09\pm0.01$ & 19/15\\
           & 7.7  & $\pi^-$ & $83\pm4$ & $-$        & $0.21\pm0.01$ & $-$           & $1.053\pm0.009$ & $0.10\pm0.01$ & 10/15\\
           & 8.8  & $\pi^-$ & $84\pm5$ & $-$        & $0.21\pm0.01$ & $-$           & $1.050\pm0.008$ & $0.10\pm0.01$ & 25/15\\
           & 12.3 & $\pi^-$ & $84\pm5$ & $-$        & $0.21\pm0.01$ & $-$           & $1.054\pm0.008$ & $0.12\pm0.01$ & 12/15\\
           & 17.3 & $\pi^+$ & $85\pm4$ & $-$        & $0.21\pm0.01$ & $-$           & $1.055\pm0.007$ & $0.13\pm0.01$ & 4/15\\
\hline
PHENIX     & 62.4 & $\pi^-$ & $84\pm4$ & $-$        & $0.22\pm0.01$ & $-$           & $1.069\pm0.008$ & $18.47\pm1.02$ &35/20\\
           &      & $\pi^+$ & $85\pm4$ & $-$        & $0.22\pm0.01$ & $-$           & $1.071\pm0.008$ & $17.60\pm1.05$ &17/20\\
           & 200  & $\pi^-$ & $80\pm4$ & $150\pm7$  & $0.20\pm0.01$ & $0.35\pm0.02$ & $1.042\pm0.009$ & $25.39\pm1.29$ &13/21\\
           &      & $\pi^+$ & $80\pm4$ & $150\pm7$  & $0.30\pm0.01$ & $0.34\pm0.02$ & $1.044\pm0.010$ & $25.77\pm1.29$ &13/21\\
\hline
STAR       & 200  & $\pi^-$ & $79\pm4$ & $150\pm7$  & $0.20\pm0.01$ & $0.35\pm0.02$ & $1.044\pm0.008$ & $0.26\pm0.01$ &20/6\\
           &      & $\pi^+$ & $79\pm4$ & $150\pm7$  & $0.20\pm0.01$ & $0.35\pm0.02$ & $1.045\pm0.008$ & $0.27\pm0.01$ &21/6\\
\hline
ALICE      & 900  & $\pi^-$ & $82\pm4$ & $170\pm8$  & $0.25\pm0.02$ & $0.36\pm0.01$ & $1.035\pm0.002$ & $1.48\pm0.08$ &74/27\\
           &      & $\pi^+$ & $83\pm4$ & $173\pm8$  & $0.25\pm0.02$ & $0.36\pm0.01$ & $1.032\pm0.002$ & $1.44\pm0.08$ &86/27\\
\hline
CMS        & 900  & $\pi^-$ & $83\pm4$ & $201\pm9$  & $0.25\pm0.02$ & $0.31\pm0.01$ & $1.032\pm0.001$ & $3.55\pm0.19$ &16/16\\
           &      & $\pi^+$ & $83\pm4$ & $200\pm9$  & $0.25\pm0.02$ & $0.31\pm0.01$ & $1.032\pm0.001$ & $3.62\pm0.19$ &21/16\\
           & 2760 & $\pi^-$ & $84\pm4$ & $202\pm9$  & $0.26\pm0.02$ & $0.38\pm0.01$ & $1.036\pm0.002$ & $4.30\pm0.23$ &53/16\\
           &      & $\pi^+$ & $84\pm4$ & $209\pm9$  & $0.26\pm0.02$ & $0.38\pm0.01$ & $1.036\pm0.002$ & $4.39\pm0.23$ &58/16\\
           & 7000 & $\pi^-$ & $85\pm4$ & $213\pm9$  & $0.28\pm0.02$ & $0.37\pm0.01$ & $1.033\pm0.003$ & $5.42\pm0.29$ &57/16\\
           &      & $\pi^+$ & $85\pm4$ & $212\pm10$ & $0.27\pm0.02$ & $0.37\pm0.01$ & $1.038\pm0.002$ & $5.48\pm0.29$ &73/16\\
           & 1300 & $\pi^-$ & $86\pm4$ & $220\pm10$ & $0.28\pm0.02$ & $0.38\pm0.01$ & $1.037\pm0.002$ & $4.96\pm0.26$ &37/16\\
           &      & $\pi^+$ & $86\pm4$ & $222\pm10$ & $0.28\pm0.02$ & $0.38\pm0.01$ & $1.038\pm0.002$ & $5.01\pm0.27$ &70/16\\
\hline
\end{tabular}%
\end{center}}
%\end{table*}
\end{sidewaystable}
\begin{multicols}{2}

\begin{figure*}
\begin{center}
\includegraphics[width=16cm]{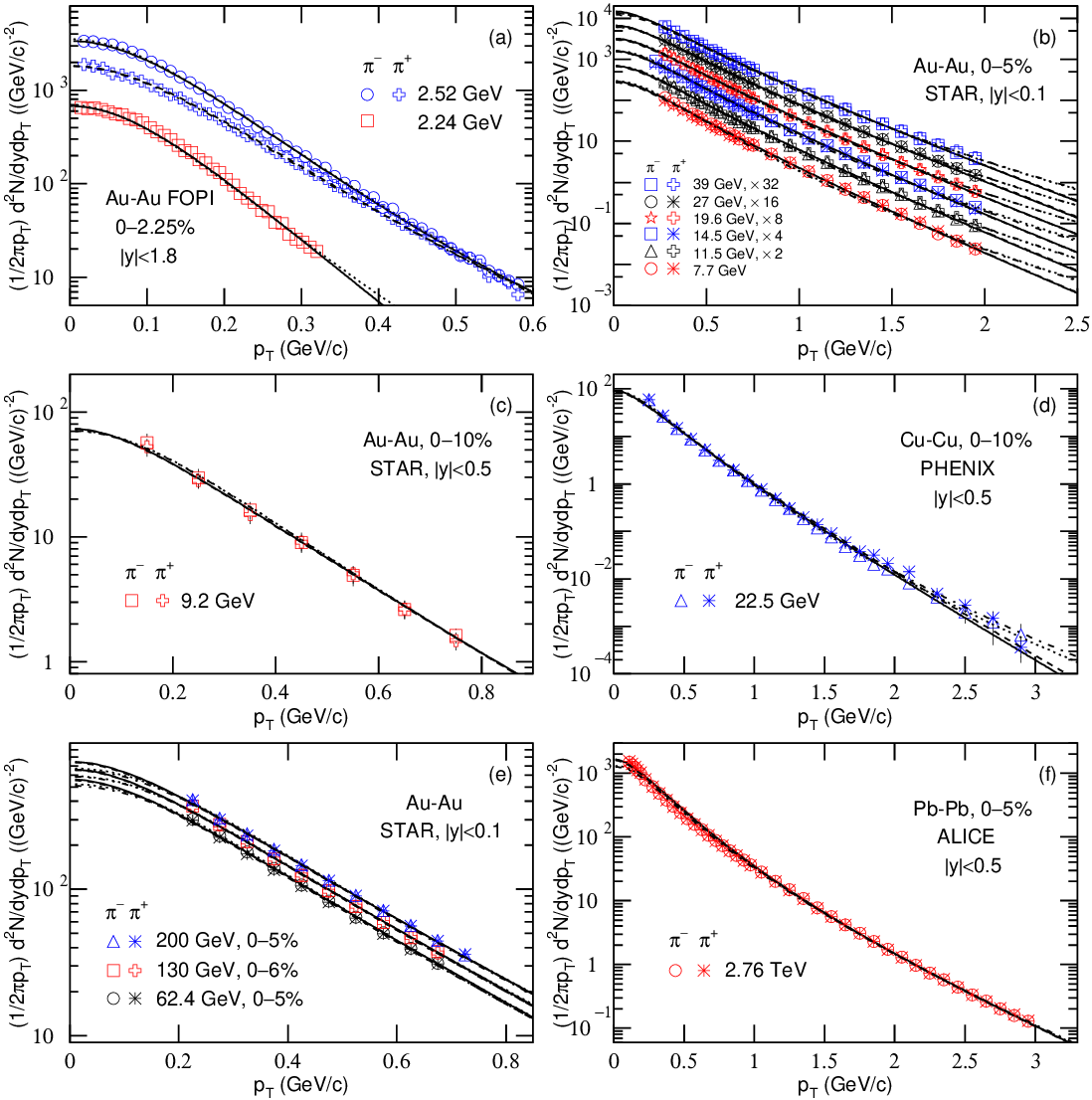}
\end{center}
{\small Fig. 2. Same as Fig. 1, but showing the results in central
Au-Au (Cu-Cu, Pb-Pb) collisions at high energies. Panels (a)--(f)
represent the data measured by the FOPI [22], STAR [27], STAR
[28], PHENIX [25], STAR [26], and ALICE [30] Collaborations,
respectively, by various symbols. The solid (dashed) curves are
our results calculated by Eqs. (1) and (3) for $\pi^-$ ($\pi^+$)
(case III), and the dotted (dot-dashed) curves are our results
calculated by Eqs. (2) and (3) for $\pi^-$ ($\pi^+$) (case IV).
Panels (a$'$)--(f$'$) and (a$''$)--(f$''$) are for ratios of the
data to fit obtained from Eqs. (1) and (3) for the case III and
from Eqs. (2) and (3) for the case IV respectively.}
\end{figure*}

\begin{figure*}
\begin{center}
\vskip -1cm
\includegraphics[width=16cm]{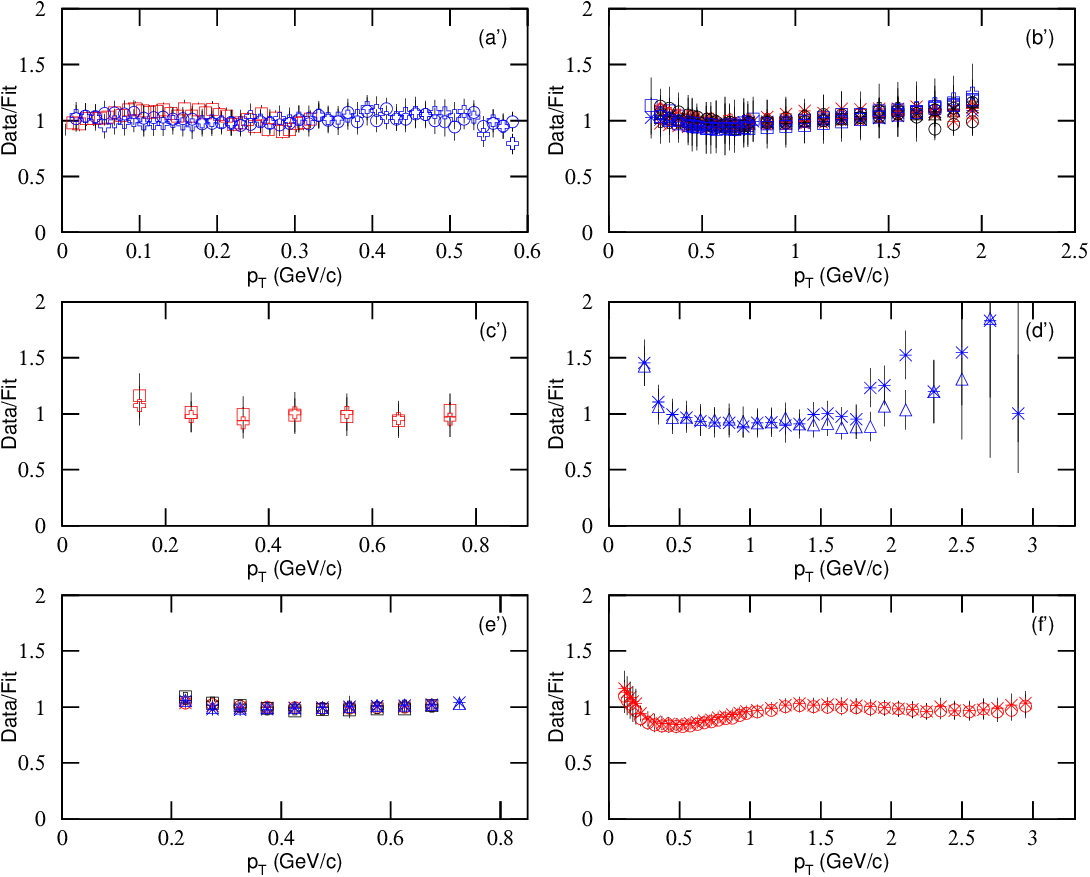}
\includegraphics[width=16cm]{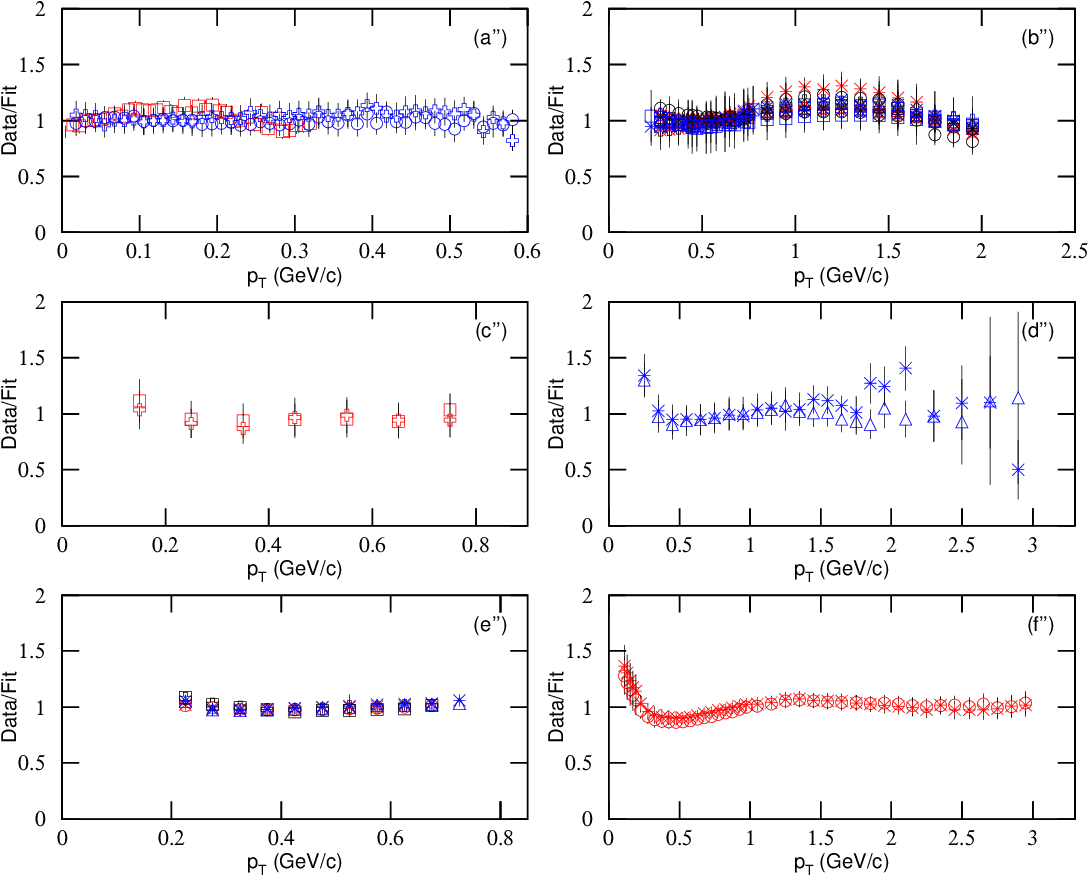}
\end{center}
{\small Fig. 2. Continued.}
\end{figure*}

\end{multicols}
\begin{sidewaystable}
%\vskip-1.cm
%\begin{table*}
{\small Table 3. Same as Table 1, but for the solid and dashed
curves in Fig. 2 in which the results at mid-rapidity in central
nucleus-nucleus collisions are presented.
\begin{center}
\begin{tabular} {ccccccccc}\\ \hline\hline Collab. and Type & $\sqrt{s_{NN}}$ (GeV) &
Particle&$T_1$ (MeV) & $T_2$ (MeV) & $\beta_{T1}$ ($c$) & $\beta_{T2}$ ($c$) & $N_0$&$\chi^2$/dof \\
\hline
FOPI Au-Au & 2.24 & $\pi^-$ & $43\pm2$ & $-$      & $0.27\pm0.01$ & $-$          & $1734.5\pm86.7$ & 37/33\\
0--2.25\%  & 2.52 & $\pi^-$ & $54\pm3$ & $98\pm5$ & $0.21\pm0.01$ & $0.42\pm0.02$& $2180.5\pm108.3$& 8/43\\
           &      & $\pi^+$ & $59\pm3$ & $98\pm5$ & $0.22\pm0.01$ & $0.43\pm0.02$& $1385.4\pm69.3$ & 5/43\\
\hline
STAR Au-Au & 7.7  & $\pi^-$ & $118\pm5$& $-$      & $0.33\pm0.02$ & $-$          & $20.1\pm1.0$    & 17/23\\
0--5\%     &      & $\pi^+$ & $118\pm5$& $-$      & $0.34\pm0.02$ & $-$          & $19.0\pm1.0$    & 27/23\\
0--10\%    & 9.2  & $\pi^-$ & $111\pm5$& $-$      & $0.34\pm0.02$ & $-$          & $22.2\pm1.1$    & 1/4\\
           &      & $\pi^+$ & $110\pm5$& $-$      & $0.33\pm0.02$ & $-$          & $21.7\pm1.1$    & 1/4\\
0--5\%     & 11.5 & $\pi^-$ & $119\pm5$& $-$      & $0.35\pm0.02$ & $-$          & $25.7\pm1.3$    & 4/23\\
           &      & $\pi^+$ & $120\pm5$& $-$      & $0.35\pm0.02$ & $-$          & $24.4\pm1.2$    & 8/23\\
           & 14.5 & $\pi^-$ & $121\pm5$& $-$      & $0.35\pm0.02$ & $-$          & $30.2\pm1.5$    & 1/25\\
           &      & $\pi^+$ & $120\pm5$& $-$      & $0.35\pm0.02$ & $-$          & $29.2\pm1.4$    & 1/25\\
           & 19.6 & $\pi^-$ & $123\pm6$& $-$      & $0.36\pm0.02$ & $-$          & $31.6\pm1.6$    & 3/23\\
           &      & $\pi^+$ & $124\pm6$& $-$      & $0.36\pm0.02$ & $-$          & $31.1\pm1.5$    & 3/23\\
           & 27   & $\pi^-$ & $124\pm6$& $-$      & $0.36\pm0.02$ & $-$          & $34.9\pm1.7$    & 3/23\\
           &      & $\pi^+$ & $124\pm6$& $-$      & $0.36\pm0.02$ & $-$          & $34.3\pm1.7$    & 3/23\\
           & 39   & $\pi^-$ & $128\pm6$& $-$      & $0.36\pm0.02$ & $-$          & $37.1\pm1.9$    & 4/23\\
           &      & $\pi^+$ & $129\pm6$& $-$      & $0.37\pm0.02$ & $-$          & $35.1\pm1.7$    & 2/23\\
           & 62.4 & $\pi^-$ & $130\pm6$& $-$      & $0.37\pm0.02$ & $-$          & $42.6\pm2.1$    & 9/4\\
           &      & $\pi^+$ & $130\pm6$& $-$      & $0.36\pm0.02$ & $-$          & $42.2\pm2.1$    & 9/4\\
0--6\%     & 130  & $\pi^-$ & $129\pm6$& $-$      & $0.37\pm0.02$ & $-$          & $50.2\pm2.5$    & 25/4\\
           &      & $\pi^+$ & $130\pm6$& $-$      & $0.38\pm0.02$ & $-$          & $49.9\pm2.5$    & 21/4\\
0--5\%     & 200  & $\pi^-$ & $132\pm6$& $-$      & $0.39\pm0.02$ & $-$          & $57.9\pm2.9$    & 6/5\\
           &      & $\pi^+$ & $131\pm6$& $-$      & $0.39\pm0.02$ & $-$          & $57.8\pm2.9$    & 9/5\\
\hline
PHENIX Cu-Cu& 22.5& $\pi^-$ & $127\pm6$& $-$      & $0.34\pm0.02$ & $-$          & $35.1\pm1.8$    & 10/20\\
0--10\%     &     & $\pi^+$ & $126\pm6$& $-$      & $0.35\pm0.02$ & $-$          & $34.9\pm1.7$    & 13/20\\
\hline
ALICE      & 2760 & $\pi^-$ & $129\pm5$&$184\pm11$& $0.43\pm0.02$ & $0.43\pm0.03$& $759.8\pm40.5$  & 37/35\\
Pb-Pb 0--5\% &    & $\pi^+$ & $129\pm5$&$184\pm14$& $0.43\pm0.02$ & $0.43\pm0.03$& $754.2\pm39.2$  & 37/35\\
\hline
\end{tabular}%
\end{center}}
%\end{table*}
\end{sidewaystable}
\begin{multicols}{2}

\end{multicols}
\begin{sidewaystable}
%\vskip-1.cm
%\begin{table*}
{\small Table 4. Same as Table 1, but one more parameter ($q$) is
added and the values correspond to the dotted and dot-dashed
curves in Fig. 2 in which the results at mid-rapidity in central
nucleus-nucleus collisions are presented.
\begin{center}
\begin{tabular} {cccccccccc}\\ \hline\hline Collab. and Type & $\sqrt{s_{NN}}$ (GeV) &
Particle&$T_1$ (MeV) & $T_2$ (MeV) & $\beta_{T1}$ ($c$) & $\beta_{T2}$ ($c$) & $q$ & $N_0$&$\chi^2$/dof \\
\hline
FOPI Au-Au & 2.24 & $\pi^-$ & $27\pm1$& $-$      & $0.16\pm0.01$ & $-$          & $1.071\pm0.014$ & $1734.5\pm86.7$ & 43/33\\
0--2.25\%  & 2.52 & $\pi^-$ & $32\pm2$& $-$      & $0.17\pm0.01$ & $-$          & $1.076\pm0.014$ & $2205.4\pm110.2$& 5/43\\
           &      & $\pi^+$ & $36\pm2$& $-$      & $0.18\pm0.01$ & $-$          & $1.078\pm0.014$ & $2205.4\pm67.6$ & 5/43\\
\hline
STAR Au-Au & 7.7  & $\pi^-$ & $75\pm4$& $-$      & $0.27\pm0.01$ & $-$          & $1.068\pm0.013$ & $20.2\pm1.0$& 63/23\\
0--5\%     &      & $\pi^+$ & $78\pm4$& $-$      & $0.27\pm0.01$ & $-$          & $1.064\pm0.013$ & $19.2\pm0.9$& 106/23\\
0--10\%    & 9.2  & $\pi^-$ & $77\pm4$& $-$      & $0.28\pm0.01$ & $-$          & $1.052\pm0.013$ & $22.7\pm1.1$& 1/4\\
           &      & $\pi^+$ & $77\pm4$& $-$      & $0.28\pm0.01$ & $-$          & $1.052\pm0.012$ & $22.5\pm1.1$& 2/4\\
0--5\%     & 11.5 & $\pi^-$ & $79\pm4$& $-$      & $0.29\pm0.01$ & $-$          & $1.065\pm0.013$ & $25.1\pm1.3$& 22/23\\
           &      & $\pi^+$ & $79\pm4$& $-$      & $0.30\pm0.01$ & $-$          & $1.065\pm0.013$ & $24.6\pm1.3$& 32/23\\
           & 14.5 & $\pi^-$ & $79\pm4$& $-$      & $0.30\pm0.01$ & $-$          & $1.068\pm0.014$ & $30.5\pm1.5$& 2/25\\
           &      & $\pi^+$ & $79\pm4$& $-$      & $0.30\pm0.01$ & $-$          & $1.068\pm0.014$ & $29.5\pm1.5$& 3/25\\
           & 19.6 & $\pi^-$ & $82\pm4$& $-$      & $0.30\pm0.02$ & $-$          & $1.066\pm0.014$ & $32.1\pm1.6$& 7/23\\
           &      & $\pi^+$ & $83\pm4$& $-$      & $0.30\pm0.02$ & $-$          & $1.067\pm0.014$ & $30.5\pm1.5$& 10/23\\
           & 27   & $\pi^-$ & $84\pm4$& $-$      & $0.32\pm0.02$ & $-$          & $1.064\pm0.014$ & $34.3\pm1.6$& 6/23\\
           &      & $\pi^+$ & $84\pm4$& $-$      & $0.32\pm0.02$ & $-$          & $1.066\pm0.014$ & $32.8\pm1.6$& 9/23\\
           & 39   & $\pi^-$ & $85\pm4$& $-$      & $0.32\pm0.02$ & $-$          & $1.065\pm0.014$ & $36.3\pm1.8$& 9/23\\
           &      & $\pi^+$ & $85\pm4$& $-$      & $0.33\pm0.02$ & $-$          & $1.068\pm0.014$ & $34.2\pm1.7$& 3/23\\
           & 62.4 & $\pi^-$ & $81\pm4$& $-$      & $0.30\pm0.02$ & $-$          & $1.078\pm0.014$ & $42.7\pm2.1$& 20/4\\
           &      & $\pi^+$ & $81\pm4$& $-$      & $0.30\pm0.02$ & $-$          & $1.080\pm0.014$ & $41.5\pm2.1$& 11/4\\
0--6\%     & 130  & $\pi^-$ & $79\pm4$& $-$      & $0.31\pm0.02$ & $-$          & $1.081\pm0.014$ & $41.5\pm2.5$& 11/4\\
           &      & $\pi^+$ & $80\pm4$& $-$      & $0.32\pm0.02$ & $-$          & $1.083\pm0.014$ & $49.5\pm2.5$& 26/4\\
0--5\%     & 200  & $\pi^-$ & $83\pm4$& $-$      & $0.32\pm0.02$ & $-$          & $1.078\pm0.014$ & $57.7\pm2.9$& 29/5\\
           &      & $\pi^+$ & $83\pm4$& $-$      & $0.33\pm0.02$ & $-$          & $1.081\pm0.014$ & $56.4\pm2.8$& 17/5\\
\hline
PHENIX Cu-Cu& 22.5& $\pi^-$ & $86\pm4$& $-$      & $0.32\pm0.02$ & $-$          & $1.054\pm0.013$ & $36.6\pm1.8$& 7/20\\
0--10\%     &     & $\pi^+$ & $86\pm4$& $-$      & $0.32\pm0.02$ & $-$          & $1.059\pm0.013$ & $35.9\pm1.8$& 15/20\\
\hline
ALICE      & 2760 & $\pi^-$ & $87\pm4$&$175\pm9$ & $0.35\pm0.02$ & $0.34\pm0.03$& $1.042\pm0.012$ &$753.5\pm37.7$& 68/35\\
Pb-Pb 0--5\% &    & $\pi^+$ & $87\pm4$&$175\pm9$ & $0.35\pm0.02$ & $0.34\pm0.03$& $1.044\pm0.012$ &$734.7\pm36.1$& 57/35\\
\hline
\end{tabular}%
\end{center}}
%\end{table*}
\end{sidewaystable}
\begin{multicols}{2}

\begin{figure*}[!htb]
\begin{center}
\includegraphics[width=16cm]{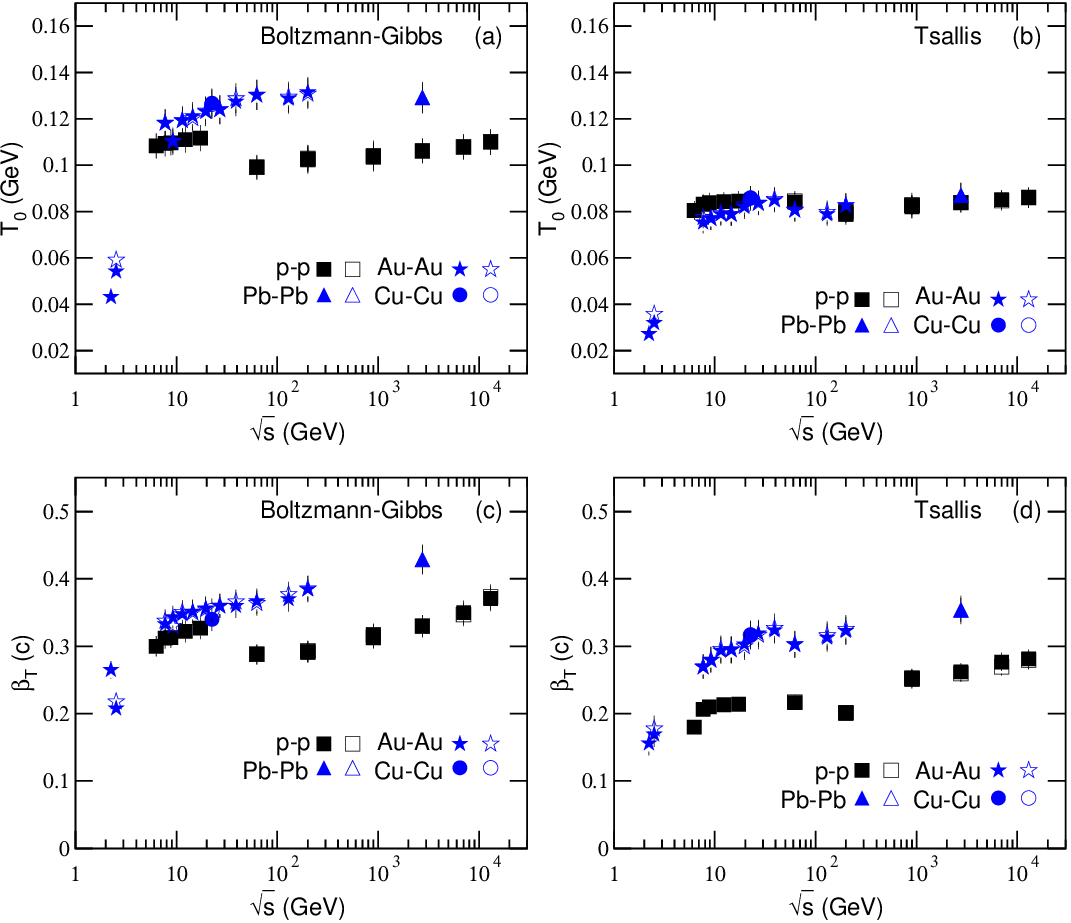}
\end{center}
{\small Fig. 3. Energy dependent (a)(b) $T_0$ and (c)(d) $\beta_T$
obtained from (a)(c) Eqs. (1) and (3) and from (b)(d) Eqs. (2) and
(3). The closed and open symbols represent the parameter values
corresponding to $\pi^-$ and $\pi^+$ respectively, which are
listed in Tables 1--4, where $\sqrt{s_{NN}}$ for nucleus-nucleus
collisions is omitted for concision.}
\end{figure*}

Figure 2 is the same as Fig. 1, but it shows the results at
mid-rapidity in central Au-Au (Cu-Cu, Pb-Pb) collisions at high
energies. The symbols in panels (a)--(f) represent the data
measured by the FOPI [22], STAR [27], STAR [28], PHENIX [25], STAR
[26], and ALICE [30] Collaborations, respectively, with different
mid-rapidity intervals, centrality intervals, and collision
energies. The solid (dashed) curves are our results calculated by
Eqs. (1) and (3) for $\pi^-$ ($\pi^+$) (case III), and the dotted
(dot-dashed) curves are our results calculated by Eqs. (2) and (3)
for $\pi^-$ ($\pi^+$) (case IV). Panels (a$'$)--(f$'$) and
(a$''$)--(f$''$) show the ratios of data to fit. The values of
various parameters, $\chi^2$, and dof corresponding to the curves
in Fig. 2 are listed in Tables 3 and 4 for the cases III and IV
respectively. One can see that Eqs. (2) and (3) describe similarly
well the $p_T$ spectra at mid-(pseudo)rapidity in central Au-Au
(Cu-Cu, Pb-Pb) collisions at different energies per nucleon pair
($\sqrt{s_{NN}}$) in center-of-mass system.

The energy dependent $T_0$ and $\beta_T$ are presented in the
upper and lower panels in Fig. 3 respectively, where the left and
right panels correspond to the results obtained from Eqs. (1) and
(3) and from Eqs. (2) and (3) respectively. The closed and open
symbols represent the parameter values corresponding to $\pi^-$
and $\pi^+$ respectively, which are listed in Tables 1 and 2 for
the cases I and II respectively, for INEL or NSD $pp$ collisions,
and in Tables 3 and 4 for the cases III and IV respectively, for
central Au-Au (Cu-Cu, Pb-Pb) collisions. One can see that the
difference between the results of $\pi^-$ and $\pi^+$ can be
neglected. In the energy dependent $T_0$ and $\beta_T$ in INEL or
NSD $pp$ collisions obtained by the blast-wave model with
Boltzmann-Gibbs statistics (case I), there are a hill at
$\sqrt{s}\approx10$ GeV, a drop at dozens of GeV, and then an
increase from dozes of GeV to above 10 TeV. The energy dependent
$T_0$ and $\beta_T$ in other three cases show a simple structure.
That is, there is a quick increase from the SIS to SPS. Then, a
slight increase or similar invariability appears from the RHIC to
LHC. Special structure is not observed around the energy bridge
from the SPS to RHIC in other three cases, though lower $T_0$ and
$\beta_T$ are obtained in the cases II and IV in which the Tsallis
statistics is used.

It should be noted that the fit of the first two points in Fig.
1(a) is to high in some cases. It means $T_0$ ($\beta_T$) for $pp$
collisions at the SPS could be higher than what is shown in Fig. 3
in the case of improving the fit. Contrarily, the fit of the first
point in Fig. 2(d) is too low. It means $T_0$ ($\beta_T$) for
central Cu-Cu collisions at 22.5 GeV could be lower than that what
is shown in Fig. 3. We give up to improve further the fit of the
first one or two points due to the trend of whole fit being worse.
Alternatively, we may use a multi-component model to improve
further the fit. The changed amounts in $T_0$ ($\beta_T$) caused
by the further improvement are small.

Since $T_0$ ($\beta_T$) is strongly related to mean transverse
momentum, $\langle p_T \rangle$, or root-mean-square transverse
momentum, $\sqrt{\langle p_T^2 \rangle}$, we plot $\langle p_T
\rangle$ against $\sqrt{s}$ ($\sqrt{s_{NN}}$) and $\sqrt{\langle
p_T^2 \rangle}$ against $\sqrt{s}$ ($\sqrt{s_{NN}}$) in the upper
and lower panels in Fig. 4 respectively, to check if the
discontinuous hill in the case I still exists, where the left and
right panels correspond to the results obtained from Eqs. (1) and
(3) and from Eqs. (2) and (3) respectively. One can see that the
hill in Fig. 3 is still existent in Fig. 4 for the case I. Similar
trend as what is shown in Fig. 3 is observed in Fig. 4 for the
other three cases.

It should be noted that the above calculations in Fig. 4 are done
according to the functions of curves in the $p_T$ range from 0 to
5 GeV/$c$, though the experimental $p_T$ ranges are different and
much shorter. If we reduce the $p_T$ range, the trend does not
change obviously due to small fraction of particles with high
$p_T$. We give up to use the experimental data to calculate
$\langle p_T \rangle$ and $\sqrt{\langle p_T^2 \rangle}$ due to
the fact that some data are incomplete in the considered $p_T$
range, in particular in the very-low-$p_T$ region.

\begin{figure*}[!htb]
\begin{center}
\includegraphics[width=16cm]{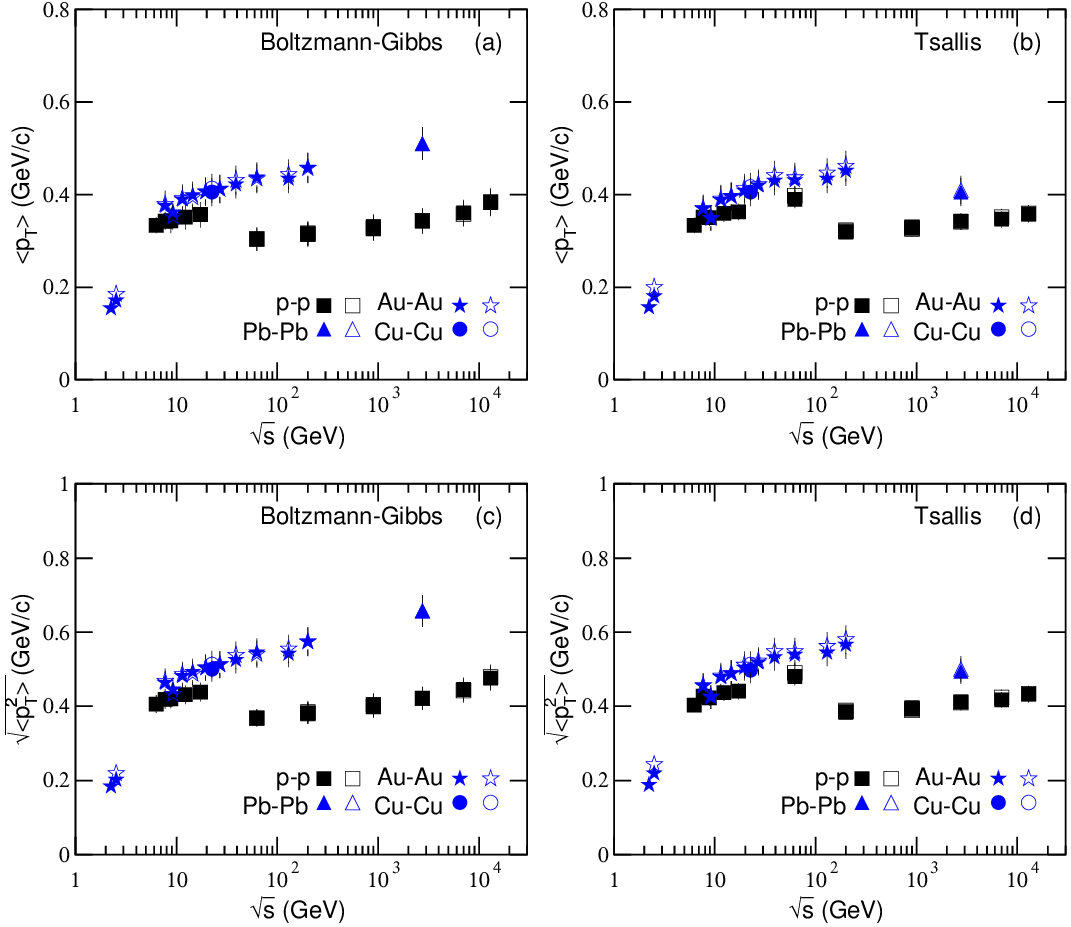}
\end{center}
{\small Fig. 4. Energy dependent (a)(b) $\langle p_T \rangle$ and
(c)(d) $\sqrt{\langle p_T^2 \rangle}$ obtained from (a)(c) Eqs.
(1) and (3) and from (b)(d) Eqs. (2) and (3). The closed and open
symbols represent the values corresponding to $\pi^-$ and $\pi^+$
respectively, where $\sqrt{s_{NN}}$ for nucleus-nucleus collisions
is omitted for concision.}
\end{figure*}

Comparing with the case I, the different situation of energy
dependence of $T_0$ ($\beta_T$) in the other three cases is caused
by larger collision systems for central nucleus-nucleus collisions
and/or one more parameter ($q$) in the Tsallis statistics. In
central nucleus-nucleus collisions, some statistical and/or
dynamical fluctuations existed in small collision systems are
smoothed because of the successional or cascade nucleon-nucleon
collisions and the multiple scattering of secondary particles. In
the Tsallis statistics, some fluctuations existed in the
Boltzmann-Gibbs statistics are smoothed due to the introduction of
$q$ which causes more entanglements in the selections of $T_0$ and
$\beta_T$. These factors cause the energy dependent $T_0$
($\beta_T$) in the other three cases to be different from that in
the case I.

In any case, the special structure around 10 GeV in the energy
dependent $T_0$ ($\beta_T$) in the case I or the different trend
which turns a corner in other three cases indicates that this
energy is special as pointed out by Cleymans [45]. The present
work shows that, in most cases, the energy dependent $T_0$
($\beta_T$) increases quickly before 10 GeV and then saturates or
increases slowly after 10 GeV with the increase of collision
energy, which also indicates this special energy (11 GeV more
specifically [45]). Some special properties result in this special
energy. The final state has the highest net baryon density at this
energy. Meanwhile, there is a transition from a baryon-dominated
to a meson-dominated final state, and the ratios of strange
particles to mesons show clear and pronounced maximums [45].

At 11 GeV, the chemical freeze-out temperature in central
nucleus-nucleus collisions is $T_{ch}\approx151$ MeV [45]. The
present work shows that, at this energy, the kinetic freeze-out
temperature in INEL or NSD $pp$ collisions is $T_0\approx115$ MeV
and that in central nucleus-nucleus collisions is $T_0\approx120$
MeV, obtained from Eqs. (1) and (3). The differences between
$T_{ch}$ and $T_0$ and between the two $T_0$ render that the
chemical freeze-out happens obviously earlier than the kinetic
freeze-out due to higher temperature at the chemical freeze-out.
The energy deposition in central nucleus-nucleus collisions is
larger than that in INEL or NSD $pp$ collisions due to higher
temperature in central nucleus-nucleus collisions. The values of
$T_0$ ($\beta_T$) obtained from Eqs. (2) and (3) are
systematically lower than those obtained from Eqs. (1) and (3) by
$\sim$20\% due to different statistics.

According to an ideal fluid consideration, the time evolution of
temperature follows $T_f=T_i(\tau_i/\tau_f)^{1/3}$, where $T_i$
($=300$ MeV) and $\tau_i$ ($=1$ fm) are the initial temperature
and proper time respectively [46, 47], and $T_f$ and $\tau_f$
denote the final temperature and time respectively. The chemical
freeze-out in central nucleus-nucleus collisions happens at about
7.8 fm due to $T_f=T_{ch}\approx151$ MeV. The kinetic freeze-out
in INEL or NSD $pp$ collisions happens at 17.8 fm due to
$T_f=T_0\approx115$ MeV, and that in central nucleus-nucleus
collisions happens at 13.8 fm due to $T_f=T_0\approx125$ MeV,
based on Eqs. (1) and (3).

Generally, in the low-$p_T$ region, $T_0$ ($\beta_T$) obtained
only from the pion spectra is less than that obtained averagely by
weighting the yields of various types of light particles. The
treatment in the present work on the extraction of $T_0$
($\beta_T$) is approximate. However, the fraction ratio of the
pion yield to total yield in high energy collisions is larger
($\sim$85\%). This approximate treatment is acceptable. The energy
dependent $T_0$ ($\beta_T$) obtained from the pion spectra is
similar to that obtained from the weighted average of the spectra
of various light particles.

From the RHIC to LHC, there are three trends for the change of
$T_0$. That is the incremental [14], decreasing [15--20], or
invariant trend [12, 13] in the energy dependent $T_0$. However,
there is only one trend for the change of $\beta_T$. That is the
incremental trend [12--20] in the energy dependent $\beta_T$. The
decreasing or invariant trend is not observed in the energy
dependent $\beta_T$. The present work shows that, in the four
cases (two collisions and two statistics), both $T_0$ and
$\beta_T$ do not decrease from the RHIC to LHC. The energy
deposition at the LHC is more than that at the RHIC.

The inconsistent results on the trend of $T_0$ ($\beta_T$) are in
fact caused by different models or methods. Similar to different
types of thermometers in thermodynamics, the standard method
extracting nuclear temperature is needed in high energy
collisions. In our opinion, the standard method should be related
to the Boltzmann-Gibbs statistics and the standard distribution
(the Boltzmann, Fermi-Dirac, and Bose-Einstein distributions)
which are the foundation of the ideal gas model in thermodynamics.
The blast-wave model with Boltzmann-Gibbs statistics [2, 3] and
the alternative method [5--11] with standard distribution are
suitable to be candidates, though other models and methods also
fit the data very well.

In the above discussions, we have used the two-component
blast-wave model in some cases. In our very recent work [48], a
superposition of the blast-wave model and an inverse power-law has
been particularly used in these cases in $pp$ collisions. It is
noted that $T_0$ ($\beta_T$) is determined by the first component
due to the fact that the spectra in low $p_T$ range is mainly
contributed by the first component. From the point of view of
extraction of $T_0$ ($\beta_T$), one can consider only the
contribution of the first component. Even the second component can
be given up in the fit process and the corresponding $p_T$ region
can be left there.

In the last part of this section, we would like to discuss further
some issues. The description of blast-wave model came from Ref.
[2] assumes the local thermal equilibrium based on a
hydrodynamical framework. From Cooper-Frye formula, one can derive
Eq. (1) for Boltzmann-Gibbs statistics. Due to the very short
interacting time, the collision system is possibly not in the
thermal equilibrium with an unified temperature, but in a few
local thermal equilibriums with different temperatures. Naturally
and in the simplest way, we can use the two-component blast-wave
model, which basically assumes that the both low-$p_T$ and
high-$p_T$ particles severally obey the above hydrodynamical
assumption.

Although the system of $pp$ collisions at low energy is probably
not in thermal equilibrium or local thermal equilibriums due to
low multiplicity, the present work treats $pp$ collisions at high
energies in which the multiplicities in most cases are not too
low. In particular, small collision systems such as $pp$ and
$p$-nucleus collisions show abundant collective behaviors [49]
which are similar to those in nucleus-nucleus collisions. These
similarities to nucleus-nucleus collisions render that the idea of
local thermal equilibrium for $pp$ collisions at high energies may
be the truth. The application of the (two-component) blast-wave
model in $pp$ collisions at high energies is acceptable.

In addition, we would like to emphasize that, comparing with
central nucleus-nucleus collisions or with the blast-wave model
with Tsallis statistics, we have observed the inconsistent trend
for the energy dependence of $T_0$ ($\beta_T$), i.e. the hill, in
$pp$ collisions analyzed by the blast-wave model with
Boltzmann-Gibbs statistics. This implies that the successional or
cascade nucleon-nucleon collisions and multiple scattering of
secondary particles in central nucleus-nucleus collisions play
important roles in the collision process. Due to these
successional collisions and multiple scattering, some statistical
and/or dynamical fluctuations in $pp$ collisions are smoothed
obviously in central nucleus-nucleus collisions. In the blast-wave
model with Tsallis statistics, the extra entropy index $q$
smoothes naturally these fluctuations.
\\

{\section{Conclusions}}

To conclude, the transverse momentum spectra of $\pi^-$ and
$\pi^+$ produced at mid-(pseudo)rapidity in INEL or NSD $pp$
collisions and in central Au-Au (Cu-Cu, Pb-Pb) collisions over an
energy range from the SIS to LHC have been analyzed by the
(two-component) blast-wave model with Boltzmann-Gibbs statistics
and with Tsallis statistics. The model results are in similarly
good agreement with the experimental data of FOPI, NA61/SHINE,
STAR, PHENIX, ALICE, and CMS Collaborations. The values of related
parameters, the kinetic freeze-out temperature $T_0$ and the
transverse flow velocity $\beta_T$, are extracted from the fit
process and the energy dependent parameters are obtained.

In INEL or NSD $pp$ collisions, and in the analysis by the
blast-wave model with Boltzmann-Gibbs statistics, both the energy
dependent $T_0$ and $\beta_T$ show a complex structure. There is a
hill at $\sqrt{s}\approx10$ GeV, a drop at dozens of GeV, and an
increase from dozes of GeV to above 10 TeV. In central Au-Au
(Cu-Cu, Pb-Pb) collisions, or in the analysis by the blast-wave
model with Tsallis statistics, both the energy dependent $T_0$ and
$\beta_T$ show simple structure. Form the SIS to LHC, there is a
quick increase before 10 GeV, and then a slight increase or
saturation after 10 GeV. The differences among the energy
dependent parameters in different cases are caused by different
collision systems and statistics. Large system smoothes the
fluctuations due to the successional collisions and multiple
scattering. The Tsallis statistics smoothes the fluctuations due
to the entropy index.

Some special properties result in the special energy of 10 GeV (11
GeV more specifically [45]). At this special energy, not only the
final state has the highest net baryon density, but also the
transition from a baryon-dominated to a meson-dominated final
state has happened. At the same time, the ratios of strange
particles to mesons show clear and pronounced maximums at this
energy. Regardless of the baryon-dominated and meson-dominated
final state, they possibly undergo a parton-dominated intermediate
state at higher energy. To search for the critical energy at which
the parton-dominated intermediate state appears initially is a
long-term target, though the critical energy is possibly around
the energy bridge from the SPS to RHIC.
\\

{\bf Data Availability}

All data are quoted from the mentioned references. As a
phenomenological work, this paper does not report new data.
\\

{\bf Conflicts of Interest}

The authors declare that there are no conflicts of interest
regarding the publication of this paper.
\\

{\bf Acknowledgments}

This work was supported by the National Natural Science Foundation
of China under Grant Nos. 11575103 and 11747319, the Shanxi
Provincial Natural Science Foundation under Grant No.
201701D121005, and the Fund for Shanxi ``1331 Project" Key
Subjects Construction.
\\

\end{multicols}

{\small
\begin{thebibliography}{99}
\setlength{\itemsep}{-1pt}

\bibitem{1}
A. Puglisi, A. Sarracino, A. Vulpiani, Phys. Rep. {\bf 709}, 1
(2017).
\bibitem{2}
E. Schnedermann, J. Sollfrank, U. Heinz, Phys. Rev. C {\bf 48},
2462 (1993)
\bibitem{3}
STAR Collaboration (B.I. Abelev {\it et al.}), Phys. Rev. C {\bf
81}, 024911 (2010).
\bibitem{4}
Z.B. Tang, Y.C. Xu, L.J. Ruan, G. van Buren, F.Q. Wang, Z.B. Xu,
Phys. Rev. C {\bf 79}, 051901(R) (2009).
\bibitem{5}
Z.B. Tang, L. Yi, L.J. Ruan, M. Shao, H.F. Chen, C. Li, B.
Mohanty, P. Sorensen, A.H. Tang, Z.B. Xu, Chin. Phys. Lett. {\bf
30}, 031201 (2013).
\bibitem{6}
K. Jiang, Y.Y. Zhu, W.T. Liu, H.F. Chen, C. Li, L.J. Ruan, Z.B.
Tang, Z.B. Xu, Phys. Rev. C {\bf 91}, 024910 (2015).
\bibitem{7}
H. Heiselberg, A.M. Levy, Phys. Rev. C {\bf 59}, 2716 (1999).
\bibitem{8}
S. Takeuchi, K. Murase, T. Hirano, P. Huovinen, Y. Nara, Phys.
Rev. C {\bf 92}, 044907 (2015).
\bibitem{9}
H.-R. Wei, F.-H. Liu, R.A. Lacey, Eur. Phys. J. A {\bf 52}, 102
(2016).
\bibitem{10}
H.-R. Wei, F.-H. Liu, R.A. Lacey, J. Phys. G {\bf 43}, 125102
(2016).
\bibitem{11}
H.-L. Lao, H.-R. Wei, F.-H. Liu, R.A. Lacey, Eur. Phys. J. A {\bf
52}, 203 (2016).
\bibitem{12}
A. Andronic, Int. J. Mod. Phys. A {\bf 29}, 1430047 (2014).
\bibitem{13}
ALICE Collaboration (B. Abelev {\it et al.}), Phys. Rev. Lett.
{\bf 109}, 252301 (2012).
\bibitem{14}
S. Zhang, Y.G. Ma, J.H. Chen, C. Zhong, Adv. High Energy Phys.
{\bf 2015}, 460590 (2015).
\bibitem{15}
S. Das for the STAR collaboration, EPJ Web of Conf. {\bf 90},
08007 (2015).
\bibitem{16}
S. Das for the STAR collaboration, Nucl. Phys. A {\bf 904--905},
891c (2013).
\bibitem{17}
S. Zhang, Y.G. Ma, J.H. Chen, C. Zhong, Adv. High Energy Phys.
{\bf 2016}, 9414239 (2016).
\bibitem{18}
STAR Collaboration (L. Adamczyk {\it et al.}), Phys. Rev. C {\bf
96}, 044904 (2017).
\bibitem{19}
X.F. Luo, Nucl. Phys. A {\bf 956}, 75 (2016).
\bibitem{20}
S. Chatterjee, S. Das, L. Kumar, D. Mishra, B. Mohanty, R. Sahoo,
N. Sharma, Adv. High Energy Phys. {\bf 2015}, 349013 (2015).
\bibitem{21}
H.-L. Lao, F.-H. Liu, B.-C. Li, M.-Y. Duan, Nucl. Sci. Tech. {\bf
29}, 82 (2018).
\bibitem{22}
FOPI Collaboration (W. Reisdorf {\it et al.}), Nucl. Phys. A {\bf
781} 459 (2007).
\bibitem{23}
NA61/SHINE Collaboration (N. Abgrall {\it et al.}), Eur. Phys. J.
C {\bf 74}, 2794 (2014).
\bibitem{24}
PHENIX Collaboration (A. Adare {\it et al.}), Phys. Rev. C {\bf
83}, 064903 (2011).
\bibitem{25}
J.T. Mitchell for the PHENIX Collaboration, PoS(CPOD2006)019,
arXiv:nucl-ex/0701079.
\bibitem{26}
STAR Collaboration (B.I. Abelev {\it et al.}), Phys. Rev. C {\bf
79}, 034909 (2009).
\bibitem{27}
STAR Collaboration (L. Adamczyk {\it et al.}), Phys. Rev. C {\bf
96}, 044904 (2017).
\bibitem{28}
STAR Collaboration (B.I. Abelev {\it et al.}), Phys. Rev. C {\bf
81}, 024911 (2010).
\bibitem{29}
ALICE Collaboration (K. Aamodt {\it et al.}), Eur. Phys. J. C {\bf
71}, 1655 (2011).
\bibitem{30}
ALICE Collaboration (B. Abelev {\it et al.}), Phys. Rev. Lett.
{\bf 109}, 252301 (2012).
\bibitem{31}
CMS Collaboration (S. Chatrchyan {\it et al.}), Eur. Phys. J. C
{\bf 72}, 2164 (2012).
\bibitem{32}
CMS Collaboration (A.M. Sirunyan {\it et al.}), Phys. Rev. D {\bf
96}, 112003 (2017).
\bibitem{33}
R. Odorico, Phys. Lett. B {\bf 118}, 151 (1982).
\bibitem{34}
UA1 Collaboration (G. Arnison {\it et al.}), Phys. Lett. B {\bf
118}, 167 (1982).
\bibitem{35}
T. Mizoguchi, M. Biyajima, N. Suzuki, Int. J. Mod. Phys. A {\bf
32}, 1750057 (2017).
\bibitem{36}
R. Hagedorn, Riv. Nuovo Cimento {\bf 6}(10), 1 (1983).
\bibitem{37}
ALICE Collaboration (B. Abelev {\it et al.}), Eur. Phys. J. C {\bf
75}, 1 (2015).
\bibitem{38}
ALICE Collaboration (K. Aamodt {\it et al.}), Phys. Lett. B {\bf
693}, 53 (2010).
\bibitem{39}
A. De Falco for the ALICE collaboration, J. Phys. G {\bf 38},
124083 (2011).
\bibitem{40}
ALICE Collaboration (B. Abelev {\it et al.}), Phys. Lett. B {\bf
710}, 557 (2012).
\bibitem{41}
HERA-B Collaboration (I. Abt {\it et al.}), Eur. Phys. J. C {\bf
50}, 315 (2007).
\bibitem{42}
ALICE Collaboration (B. Abelev {\it et al.}), Phys. Lett. B {\bf
718}, 295 (2012) and Erratum: Phys. Lett. B {\bf 748}, 472 (2015).
\bibitem{43}
I. Lakomov for the ALICE collaboration, Nucl. Phys. A {\bf 931},
1179 (2014).
\bibitem{44}
ALICE Collaboration (B. Abelev {\it et al.}), Phys. Lett. B {\bf
708}, 265 (2012).
\bibitem{45}
J. Cleymans, arXiv:1711.02882 [hep-ph] (2017).
\bibitem{46}
J.D. Bjorken, Phys. Rev. D {\bf 27}, 140 (1983).
\bibitem{47}
K. Okamoto, C. Nonaka, Eur. Phys. J. C {\bf 77}, 383 (2017).
\bibitem{48}
L.-L. Li, F.-H. Liu, Physics {\bf 2}, 277 (2020).
\bibitem{49}
H.C. Song, Y. Zhou, K. Gajdo{\v s}ov{\' a}, Nucl. Sci. Tech. {\bf
28}, 99 (2017).

\end{thebibliography}}
\end{document}